\documentclass[12pt,onecolumn,floatfix,amsmath,nofootinbib,amssymb,floatfix,aps,prd,showpacs]{revtex4}
\usepackage{slashed}
\usepackage[level]{datetime}
\usepackage{graphicx,color,dcolumn,booktabs,bm}
\usepackage{longtable,lscape}
\usepackage{txfonts}
\usepackage{subfigure}
\usepackage{float}
\usepackage{makecell}
\usepackage{tabularx}
\usepackage{array}
\usepackage{threeparttable}
\usepackage{ragged2e}
\newcolumntype{P}[1]{>{\centering\hspace{0pt}}p{#1}}
\newcolumntype{Z}{>{\centering\let\newline\\\arraybackslash\hspace{0pt}}x}

\usepackage{booktabs}
\usepackage{overpic}
\usepackage{amssymb}
\usepackage{indentfirst}
\usepackage{feynmf}   
\usepackage{slashed}  
\usepackage{cases}
\usepackage{color}
\usepackage{multirow}
\usepackage{epstopdf}
\usepackage{longtable}
\usepackage{indentfirst}
\usepackage{graphicx,color,dcolumn,booktabs,bm}
\usepackage[colorlinks,
            citecolor=blue,
            anchorcolor=red,
            menucolor=red,
            linkcolor=red,
            filecolor=red,
            runcolor=red,
            urlcolor=blue,
            frenchlinks=red]{hyperref}
\graphicspath{{Figures/}} %
\allowdisplaybreaks
\usepackage{CJK}
\begin{document}
\title{Twist-2 distribution amplitudes of $a_{0}(980)$ and $a_{0}(1450)$}
\author{Wei, Hong$^{1,2}$}
\email{1540862997@qq.com}
\author{Di Gao$^{2,3}$}
\email{digao@impcas.ac.cn}
\author{Yanjun Sun$^{2,3,4}$}
\email{sunyanjun@nwnu.edu.cn}

\affiliation{ $^1$Basic Teaching Office, Shaanxi Fashion Engineering University,  Xianyang Shaanxi, 712046, China}
\affiliation{ $^2$Institute of Theoretical Physics, College of Physics and Electronic Engineering, Northwest Normal University,  Lanzhou 730070, China}
\affiliation{ $^3$Institute of Modern Physics, Chinese Academy of Sciences, Lanzhou 730000, China}
\affiliation{ $^4$Lanzhou Center for Theoretical Physics, Lanzhou University, Lanzhou 730070, China}


\begin{abstract}
We investigate the twist-2 distribution amplitudes of the scalar mesons $a_{0}(980)$ and $a_{0}(1450)$ in the two-quark picture. The moments of these scalar mesons are obtained up to the third order with QCD sum rules method. With these moments, the first two Gegenbauer coefficients are determined and utilized to analyze the twist-2 distribution amplitudes. Our numerical results indicate that the meson $a_{0}(980)$ favors a conventional two-quark ground state. The paper concludes with an examination of the form factors for the transitions $B/D\rightarrow a_{0}$.
\end{abstract}

\maketitle

\section{Introduction}
In the early 1990s, two scalar mesons with properties that challenged the quark model were discovered \cite{Lepage:1980fj,Chernyak:1981zz}: the mass-degenerate and relatively narrow resonances $f_{0}(980)$ and $a_{0}(980)$. Cribov and colleagues proposed that they could be novel mesons, suggesting that their quark-antiquark configurations are $\frac{u\bar{u}+d\bar{d}}{\sqrt{2}}$ for the isoscalar $f_{0}(980)$ and $\frac{u\bar{u}-d\bar{d}}{\sqrt{2}}$ for the isovector $a_{0}(980)$ \cite{Amsler:1995tu}. Meanwhile, the mass and width patterns of those scalar mesons led R. L. Jaffe to propose tetraquark assignments for a complete light scalar nonet, potentially explaining the mass degeneracy of $f_{0}(980)$ and $a_{0}(980)$ due to identical four-quark compositions \cite{Jaffe:1976ig}. Despite these scalar mesons, $f_{0}(600)$, $f_{0}(500)$, $K_{0}^{*}(700)$, $f_{0}(980)$ and $a_{0}(980)$, currently forming a complete light nonet in Particle Data Group (PDG) listings, there remains a diversity of perspectives on $a_{0}(980)$ \cite{BaBar:2009rrj}.
\par
Over the past two decades, significant advancements have been made in both theoretical \cite{Wu:2010zc,Szczepaniak:1993kk,Huang:1983fw,Chernyak:1983ej,Cao:1997st,Wei:2003fm} and experimental studies \cite{E852:1996san,CDF:1997chq,Bediaga:2002ei,BaBar:2001phq,KLOE:2005wnj,Belle:2006wcd} of the scalar mesons $a_0(980)$ and $a_0(1450)$. Researchers have employed various methods, including perturbative QCD approaches \cite{Pennington:2002nb,Brambilla:2019esw,Chiba:2023ftg,Chiba:2024cny}, QCD sum rules \cite{Latorre:1985uy,Surguladze:1990sp,Gokalp:2001ku,Markushin:2000ff,Steele:1999bg,Shi:1999fx,Elias:1998bq,Choe:1996uc,Liu:1993rba,Lu:2006fr}, light-cone sum rules \cite{Gokalp:2003ih,Aliev:2001mm,Wang:2010zg,Aydin:2006eh,Aydin:2005gw,Gokalp:2004ny,DeFazio:2010xs}, and phenomenological models \cite{Mennessier:2010xz,Belyaev:2010dz,Narison:2021xhc}, to investigate these particles, focusing on aspects such as distribution amplitudes. 
\par
Distribution amplitudes are fundamental tools in hadronic physics, essential for understanding the internal momentum distribution of quarks and gluons within hadrons. They serve as crucial inputs in theoretical frameworks such as light-cone sum rules and QCD factorization, which govern hard scattering amplitudes in various reactions. The normalization of the distribution amplitudes of a meson is the decay constant. These comprehensive studies enhance our understanding of hadronic structures and the dynamics governing meson decays, offering valuable information for experimental validation at facilities like LHCb and Belle \uppercase\expandafter{\romannumeral2}.
\par
In 2010, to figure out the physical properties of the scalar meson $a_{0}(980)$, Ref. \cite{Wang:2009azc} introduced a model-independent approach to probe the quark structure of light scalar mesons: they defined a critical ratio parameter $R$ as the ratio of partial decay widths, which crucially differs between configurations: $R=1$ for a two-quark ($q\bar{q}$) configuration and $R=3$ for a four-quark ($qq\bar{q}\bar{q}$) structure. These predictions have been experimentally validated by the Beijing Spectrometer (BES \uppercase\expandafter{\romannumeral3}) collaboration \cite{BESIII:2018sjg} and BES \uppercase\expandafter{\romannumeral3} reports that $R=2.03\pm0.95\pm0.06$. In a sense, semileptonic decays related to $D$-meson inherently offer a clean experimental platform to probe the $a_{0}(980)$ resonance, which dynamically emerges from an isovector configuration of $u/d$-quark-antiquark pairs. Therefore, we construct the distribution amplitudes of $a_{0}$ within the two-quark picture in this work.
\par
In this work, we first apply QCD sum rules to derive expressions for the moments of the twist-2 distribution amplitudes of light scalar mesons $a_{0}(980)$ and $a_{0}(1450)$, analyze their characteristics and provide the first two non-zero moments. Subsequently, we examine the dependence of these moments on the Borel parameter $M^2$ and the threshold $s_0$. Finally, we present the transition form factors for $B/D\rightarrow a_{0}(980), a_{0}(1450)$ decay processes by light cone sum rule.
\par
This paper is organized as follows: In Sec. \uppercase\expandafter{\romannumeral2}, we derive the general sum rules of moments for twist-2 distribution amplitudes of scalar mesons. In Sec. \uppercase\expandafter{\romannumeral3}, we provide the numerical results for the first two moments for $a_{0}(980)$ and $a_{0}(1450)$. The form factors of $B/D\rightarrow a_{0}$ are also obtained. Sec. \uppercase\expandafter{\romannumeral4} presents our conclusions.
\section{QCD sum rule analyses}
\subsection{Twist-2 distribution amplitudes of scalar mesons}
The twist-2 distribution amplitude $\phi_{s}(x,\mu)$ of scalar mesons $a_{0}$ in the two-quark picture is characterized by the following definition \cite{Huang:2004tp}
\begin{equation}
\langle S(p) |\bar{q}_{2}(z_{2})\gamma_{\mu}q_{1}(z_{1})|0\rangle=p_{\mu}\int^{1}_{0}dxe^{i(xpz_{2}+\bar{x}pz_{1})}\phi_{s}(x,\mu),
\label{doda}
\end{equation}
where $x$($\bar{x}$) represents the momentum fraction of the quark $q_{2}$($q_{1}$) within the scalar meson, with $\bar{x}=1-x$, $p$ is the momentum of scalar meson and $\mu$ is the energy scale. The distribution amplitude subjects to the normalization condition
\begin{equation}
\int^{1}_{0}dx\phi_{s}(x,\mu)=f_{s},
\end{equation}
where $f_{s}$ is the decay constant of scalar meson.
\par
For a light scalar meson, depicted within a two-parton framework, it is capable of coupling to both vector and scalar quark current operators. Consequently, the decay constants can be defined as \cite{Cheng:2005nb}
\begin{equation}
\begin{aligned}
\langle S(p)|\bar{q}_{2}\gamma_{\mu}q_{1}|0\rangle&=f_{s}p_{\mu},\\
\langle S(p)|\bar{q}_{2}q_{1}|0\rangle&=m_{s}\bar{f_{s}} .\\ \notag
\end{aligned}
\end{equation}
For a charged scalar meson, the decay constants  $f_{s}$ and $\bar{f}_{s}$ are related through the equation of motion
\begin{equation}
\begin{aligned}
\mu_{s}f_{s}&=\bar{f}_{s},\\
\mu_{s}&=\frac{m_{s}}{m_{2}(\mu)-m_{1}(\mu)},\\ \notag
\end{aligned}
\end{equation}
where $m_{1}$, $m_{2}$ and $m_{s}$ denote the masses of $q_{1}$, $q_{2}$ and $a_{0}$, respectively.
\par
Based on the conformal symmetry hidden in the QCD Lagrangian, the distribution amplitude can be expanded in a series of Gegenbauer polynomial $C_{m}^{3/2}$ with increasing conformal spin as \cite{Yang:2007zt}
\begin{equation}
\phi_{s}(x,\mu)=\bar{f_{s}}6x(1-x)[B_{0}+\mu_{s}\sum_{m=1}^{\infty}B_{m}(\mu)C_{m}^{\frac{3}{2}}(2x-1)].
\label{phi-G}
\end{equation}
Here, $C_{m}^{\frac{3}{2}}$ represents the Gegenbauer polynomials, and $B_{m}$ is the Gegenbauer coefficients.
\par
Utilizing the background field method, we can compute the moment of the twist-2 distribution amplitude defined in Equation (\ref{doda}). From Equation (\ref{doda}), it is straightforward to deduce
\begin{equation}
\langle0|\bar{q}_{2}\slashed{z}(iz\cdot\mathop{D}\limits^{\leftrightarrow}q_{1})^{n}|S(p)\rangle=(z\cdot p)^{n+1}\bar{f}_{s}\langle\xi^{n}\rangle,
\end{equation}
with the moment
\begin{equation}
\langle\xi^{n}\rangle=\int_{0}^{1}dx(2x-1)^{n}\phi_{s}(x,\mu).
\end{equation}
To calculate the above $\langle\xi^{n}\rangle$ , we consider the following correlation function
\begin{equation}
\begin{aligned}
\Pi_{n}(z,q)&=(z\cdot q)^{n+1}I_{n}(q^{2})\\
&=i\int d^{4}xe^{iqx}\langle0|T\{O_{n}(x)O^{+}(0)\}|0\rangle,
\end{aligned}
\end{equation}
with
\begin{equation}
O_{n}=\bar{q}_{2}\slashed{z}(iz\cdot\mathop{D}\limits^{\leftrightarrow})^{n}q_{1}; \\
O=\bar{q}_{2}q_{1}.
\end{equation}
\par
\par
In the deep Euclidean region, the correlation function is calculated using the operator product expansion (OPE), truncated at dimension-6 operators, and presented as follows:
\begin{equation}
\begin{aligned}
I_{n}(q^{2})&=\frac{3}{8\pi^{2}}[5+3\times(-1)^{n}+2n]\frac{1}{(n+1)(n+2)}\ln{\frac{-q^{2}}{\mu^{2}}}[(-1)^{n+1}m_{1}+m_{2}]\\
&-\frac{2n+1}{(2n+2)q^{2}}[\langle0|\bar{q}_{2}q_{2}|0\rangle+(-1)^{n+1}\langle0|\bar{q}_{1}q_{1}|0\rangle]
+\frac{2n+1}{4q^{4}}[m_{1}^{2}\langle0|\bar{q}_{2}q_{2}|0\rangle+(-1)^{n+1}m_{2}^{2}\langle0|\bar{q}_{1}q_{1}|0\rangle]\\
&+\frac{m_{1}m_{2}}{4q^{4}}[\langle0|\bar{q}_{2}q_{2}|0\rangle+(-1)^{n+1}\langle0|\bar{q}_{1}q_{1}|0\rangle]
-\frac{10+n}{24}\frac{1}{q^{4}}[\langle0|g\bar{q}_{2}TG\sigma q_{2}|0\rangle+(-1)^{n+1}\langle0|g\bar{q}_{1}TG\sigma q_{1}|0\rangle]\\
&-\frac{16(n+1)\pi}{81}\frac{1}{q^{6}}[m_{1}\langle0|\sqrt{\alpha_{s}}\bar{q}_{2}q_{2}|0\rangle^{2}+(-1)^{n+1}m_{2}\langle0|\sqrt{\alpha_{s}}\bar{q}_{1}q_{1}|0\rangle^{2}].
\label{In}
\end{aligned}
\end{equation}
\par
The correlation function may also be evaluated at the hadron level. This involves inserting a complete set of states with quantum numbers matching those of the current operator $O$ into the two-point correlation function. We arrive at the following hadronic representation of the correlation function
\begin{equation}
{\rm{Im}}I_{n}^{had}=\pi\delta(s-m_{s}^{2})m_{s}\bar{f}_{s}^{2}\langle\xi^{n}\rangle+\pi\rho^{h}_{s}\theta(s-s_{0}),
\label{Im}
\end{equation}
where $s_{0}$ is the threshold between ground and excited states. The second term needs to be dealt with. Further, employing the dispersion relation, the correlation function at the hadronic level is matched with the OPE side as follows \cite{Colangelo:2000dp}:
\begin{equation}
\frac{1}{\pi}\int_{0}^{\infty}ds\frac{{\rm{Im}}I_{n}(s)}{s-q^{2}}=I^{QCD}_{n}(q^{2}).
\end{equation}
Applying quark-hadron duality to Equations~(\ref{In}) and~(\ref{Im}), one can obtain
\begin{equation}
{\rm{Im}}I_{n}^{had}=\pi\delta(s-m_{s}^{2})m_{s}\bar{f}_{s}^{2}\langle\xi^{n}\rangle+\frac{3}{8\pi^{2}}[5+3\times(-1)^{n}+2n]\frac{1}{(n+2)(n+1)}\theta(s-s_{0}).
\end{equation}
Matching the hadronic and OPE sides, we derive the sum rule
\begin{equation}
I_{n}(q^{2})=\pi\delta(s-m_{s}^{2})m_{s}\bar{f}_{s}^{2}\langle\xi^{n}\rangle+\frac{3}{8\pi^{2}}[5+3\times(-1)^{n}+2n]\frac{1}{(n+2)(n+1)}\theta(s-s_{0}).
\label{In=xi}
\end{equation}
To suppress the contribution of higher states, represented by the second term in Equation (\ref{Im}), we apply the Borel transformation to both sides of the equation. This yields the following expression:
\begin{equation}
\frac{1}{\pi}\frac{1}{M^{2}}\int^{\infty}_{0} ds e^{-\frac{s}{M^{2}}}{\rm{Im}}I_{n}(s)=B_{M^2}I^{QCD}_{n}(q^{2}),
\end{equation}
where the Borel transform is defined as
\begin{equation}
B_{M^2}=\lim_{Q^{2},n\rightarrow \infty,\frac{Q^{2}}{n}=M^{2}}\frac{1}{(n-1)!}(Q^{2})^{n}(-\frac{d}{dQ^{2}})^{n},
\label{Bm}
\end{equation}
here, $M^{2}$ is Borel parameter.
\par
Finally, comparing Equations (\ref{In}) and (\ref{In=xi}), the expression for the moment $\langle\xi^{n}\rangle$ is derived after applying the Borel transformation (\ref{Bm}) to hadron and OPE sides as follows:
\begin{equation}
\begin{aligned}
\langle\xi^{n}\rangle&=\frac{e^{\frac{m^{2}_{s}}{M^{2}}}}{m_{s}\bar{f}^{2}_{s}}\{\frac{3}{16\pi^{2}}[3+(-1)^{n}+2n]\frac{1}{(n+2)(n+1)}[(-1)^{n+1}m_{1}+m_{2}](1+e^{-\frac{s_{0}}{M^{2}}})\\
&+\langle0|\bar{q}_{2}q_{2}|0\rangle+(-1)^{n+1}\langle0|\bar{q}_{1}q_{1}|0\rangle
+\frac{2n+1}{2M^{2}}[m_{1}^{2}\langle0|\bar{q}_{2}q_{2}|0\rangle+(-1)^{n+1}m_{2}^{2}\langle0|\bar{q}_{1}q_{1}|0\rangle]\\
&+\frac{m_{1}m_{2}}{2M^{2}}[\langle0|\bar{q}_{2}q_{2}|0\rangle+(-1)^{n+1}\langle0|\bar{q}_{1}q_{1}|0\rangle]
-\frac{2n}{3}\frac{1}{M^{2}}[\langle0|g\bar{q}_{2}TG\sigma q_{2}|0\rangle+(-1)^{n+1}\langle0|g\bar{q}_{1}TG\sigma q_{1}|0\rangle]\\
&+\frac{8n\pi}{81}\frac{1}{M^{4}}[m_{1}\langle0|\sqrt{\alpha_{s}}\bar{q}_{2}q_{2}|0\rangle^{2}+(-1)^{n+1}m_{2}\langle0|\sqrt{\alpha_{s}}\bar{q}_{1}q_{1}|0\rangle^{2}]
\}.
\end{aligned}
\end{equation}
Since the contributions from higher dimensions are minimal, we truncate the OPE expansion at dimension 6 here. Meanwhile, for scalar mesons such as the $a_{0}(980)$ and $a_{0}(1450)$, where the quark-antiquark pair is $\bar{u}d$, the aforementioned equation is reformulated as follows
\begin{equation}
\begin{aligned}
\langle\xi^{n}\rangle&=\frac{e^{\frac{m^{2}_{s}}{M^{2}}}}{m_{s}\bar{f}^{2}_{s}}\{\frac{3}{16\pi^{2}}[3+(-1)^{n}+2n]\frac{1}{(n+2)(n+1)}[(-1)^{n+1}m_{u}+m_{d}](1+e^{-\frac{s_{0}}{M^{2}}})\\
&+\langle0|\bar{d}d|0\rangle+(-1)^{n+1}\langle0|\bar{u}u|0\rangle
+\frac{2n+1}{2M^{2}}[m_{u}^{2}\langle0|\bar{d}d|0\rangle+(-1)^{n+1}m_{d}^{2}\langle0|\bar{u}u|0\rangle]\\
&+\frac{m_{u}m_{d}}{2M^{2}}[\langle0|\bar{d}d|0\rangle+(-1)^{n+1}\langle0|\bar{u}u|0\rangle]
-\frac{2n}{3}\frac{1}{M^{2}}[\langle0|g\bar{d}TG\sigma d|0\rangle+(-1)^{n+1}\langle0|g\bar{u}TG\sigma u|0\rangle]\\
&+\frac{8n\pi}{81}\frac{1}{M^{4}}[m_{u}\langle0|\sqrt{\alpha_{s}}\bar{d}d|0\rangle^{2}+(-1)^{n+1}m_{d}\langle0|\sqrt{\alpha_{s}}\bar{u}u|0\rangle^{2}]
\}.
\label{xin}
\end{aligned}
\end{equation}
The analysis of Equation~(\ref{xin}) reveals that the contribution from $\langle\xi^{n}\rangle$ is minimal for even values of $n$. Consequently, our focus shifts to the non-vanishing odd moments, specifically $\langle\xi^{1}\rangle$ and $\langle\xi^{3}\rangle$, which are given by
\begin{equation}
\begin{aligned}
\langle\xi^{1}\rangle&=\frac{e^{\frac{m^{2}_{s}}{M^{2}}}}{m_{s}\bar{f}^{2}_{s}}\{-\frac{1}{8\pi}[m_{u}+m_{d}](1+e^{-\frac{s_{0}}{M^{2}}})\\
&+\langle0|\bar{d}d|0\rangle+\langle0|\bar{u}u|0\rangle
+\frac{3}{2M^{2}}[m_{u}^{2}\langle0|\bar{d}d|0\rangle+m_{d}^{2}\langle0|\bar{u}u|0\rangle]\\
&+\frac{m_{u}m_{d}}{2M^{2}}[\langle0|\bar{d}d|0\rangle+\langle0|\bar{u}u|0\rangle]
-\frac{2}{3}\frac{1}{M^{2}}[\langle0|g\bar{d}TG\sigma d|0\rangle+\langle0|g\bar{u}TG\sigma u|0\rangle]\\
&+\frac{8\pi}{81}\frac{1}{M^{4}}[m_{u}\langle0|\sqrt{\alpha_{s}}\bar{d}d|0\rangle^{2}+m_{d}\langle0|\sqrt{\alpha_{s}}\bar{u}u|0\rangle^{2}]
\},
\end{aligned}
\end{equation}

\begin{equation}
\begin{aligned}
\langle\xi^{3}\rangle&=\frac{e^{\frac{m^{2}_{s}}{M^{2}}}}{m_{s}\bar{f}^{2}_{s}}\{-\frac{3}{40\pi}[m_{u}+m_{d}](1+e^{-\frac{s_{0}}{M^{2}}})\\
&+\langle0|\bar{d}d|0\rangle+\langle0|\bar{u}u|0\rangle
+\frac{5}{2M^{2}}[m_{u}^{2}\langle0|\bar{d}d|0\rangle+m_{d}^{2}\langle0|\bar{u}u|0\rangle]\\
&+\frac{m_{u}m_{d}}{2M^{2}}[\langle0|\bar{d}d|0\rangle+\langle0|\bar{u}u|0\rangle]
-\frac{2}{M^{2}}[\langle0|g\bar{d}TG\sigma d|0\rangle+\langle0|g\bar{u}TG\sigma u|0\rangle]\\
&+\frac{8\pi}{27}\frac{1}{M^{4}}[m_{u}\langle0|\sqrt{\alpha_{s}}\bar{d}d|0\rangle^{2}+m_{d}\langle0|\sqrt{\alpha_{s}}\bar{u}u|0\rangle^{2}]
\}.
\end{aligned}
\end{equation}
The renormalization group equations for the decay constant, quark mass, and condensate terms are expressed as follows \cite{Yang:1993bp}
\begin{equation}
\begin{aligned}
\bar{f}_{s}(\mu_{0})&=\bar{f}_{s}(\mu)\Big(\frac{\alpha_{s}(\mu)}{\alpha_{s}(\mu_{0})}\Big)^{\frac{4}{b}},\\
m_{q;\mu_{0}}&=m_{q;\mu}\Big(\frac{\alpha_{s}(\mu)}{\alpha_{s}(\mu_{0})}\Big)^{\frac{4}{b}},\\
\langle0|\bar{q}q|0\rangle_{|\mu=\mu_{0}}&=\langle0|\bar{q}q|0\rangle_{\mu}\Big(\frac{\alpha_{s}(\mu)}{\alpha_{s}(\mu_{0})}\Big)^{\frac{4}{b}},\\
\langle0|g\bar{q}TG\sigma q|0\rangle_{|\mu=\mu_{0}}&=\langle0|g\bar{q}TG\sigma q|0\rangle_{\mu}\Big(\frac{\alpha_{s}(\mu)}{\alpha_{s}(\mu_{0})}\Big)^{\frac{4}{b}},\\
\langle0|\sqrt{\alpha_{s}}\bar{q}q|0\rangle^{2}_{|\mu =\mu_{0}}&=\langle0|\alpha_{s}\bar{q}q|0\rangle^{2}_{\mu}\Big(\frac{\alpha_{s}(\mu)}{\alpha_{s}(\mu_{0})}\Big)^{\frac{4}{b}}.\\
\end{aligned}
\end{equation}
Here, $b=\frac{33-2n_{f}}{3}$, where $n_{f}$ represents the number of quark flavors and $\mu_{0}$ represents a known energy scale. Additionally, the orthogonality relation for the Gegenbauer polynomials is given by
\begin{equation}
\int^{1}_{0}dx x(1-x)C^{\frac{3}{2}}_{m}(2x-1)C^{\frac{3}{2}}_{n}(2x-1)=\frac{(n+1)(n+2)}{4(2n+3)}\delta_{mn}.
\end{equation}
Considering Equation (\ref{phi-G}), we can obtain the following relations between the Gegenbauer coefficients $B_{m}$ and the moments $\langle\xi^{n}\rangle$
\begin{equation}
\begin{aligned}
B_{0}&=\langle\xi^0\rangle,\\
B_{1}&=\frac{5}{3}\langle\xi^{1}\rangle,\\
B_{3}&=\frac{21}{4}\langle\xi^{3}\rangle-\frac{9}{4}\langle\xi^{1}\rangle,\\
&…
\label{B-xi}
\end{aligned}
\end{equation}
The renormalization group equations governing the Gegenbauer moments are articulated as
\begin{equation}
B_{n}(\mu)=B_{n}(\mu_{0})\Big(\frac{\alpha_{s}(\mu_{0})}{\alpha_{s}(\mu)}\Big)^{-\frac{\gamma_{n}-4}{b}},
\end{equation}
where the anomalous dimension $\gamma_{n}$ is
\begin{equation}
\gamma_{n}=C_{f}\Big(1-\frac{2}{(n+1)(n+2)}\Big)+\sum_{j=2}^{n+1}\frac{1}{j}.
\end{equation}
The constant $C_{f}$ is assigned the value of $\frac{4}{3}$, which is pivotal in our subsequent calculations.
\par
\subsection{The transition form factors for $B/D\rightarrow S$ with the light-cone sum rules}

To elucidate the form factors, we construct the two-point correlation function as follows \cite{Brodsky:1981rp,Geshkenbein:1982zs,Radyushkin:2009zg,Arleo:2010yg}
\begin{equation}
\Pi_{\mu}(p,q)=i\int d^{4}x'e^{iqx'}\langle S(p)|T\{\bar{q}_{2}(x')\gamma_{\mu}(1+\gamma_{5})Q(x'),\bar{Q}(0)i(1+\gamma_{5})q_{1}(0)\}|0\rangle,\\
\end{equation}
where $S$ denotes the light scalar mesons, and $Q$ represents either $b$ or $c$ quark. 
\par
Typically, the correlation function can be expressed from two distinct viewpoints: \textbf{1}) at the hadronic level, by inserting a complete set of meson states between the two currents, and \textbf{2}) at the quark and gluon level, using OPE for the correlation function. 
\par
On the hadronic level, the correlation function is read
\begin{equation}
\Pi_{\mu}(p,q)=\frac{\langle S(p)|\bar{q}_{2}\gamma_{\mu}\gamma_{5}Q|M\rangle \langle M|\bar{Q}i\gamma_{5}q_{1}|0\rangle}{m^{2}_{B/D}-(p+q)^{2}}+higher \quad states,\\
\end{equation}
where $|M\rangle$ represents the pseudoscalar meson $B/D$. The matrix element is parameterized by form factors as
\begin{equation}
\begin{aligned}
\langle S(p)|\bar{q}_{2}\gamma_{\mu}\gamma_{5}Q|M\rangle=-2if_{+}(q^{2})p_{\mu}-i[f_{+}(q^{2}+f_{-}(q^{2}))]q_{\mu},\\
\end{aligned}
\end{equation}
with $f_{+}$ and $f_{-}$ being the transition form factors for $B/D\rightarrow S$, $m_{B/D}$ representing the masses of the $B/D$ meson. Subsequently, the hadronic side is expressed as
\begin{equation}
\Pi_{\mu}(p,q)=-\frac{2f_{+}(q^{2})p_{\mu}+[f_{+}(q^{2})+f_{-}(q^{2})]q_{\mu}}{m_{B/D}^{2}-(p+q)^{2}}\frac{m^{2}_{B/D}f_{B/D}}{m_{Q}+m_{q_{1}}}+\int_{s_{0}}^{\infty}\frac{\rho(s)ds}{s-(p+q)^{2}},\\
\end{equation}
where $f_{B/D}$ is the decay constant of $B/D$ meson, $m_{Q}$ and $m_{q_{1}}$ represent the masses of heavy quark $Q$ and light quark $q_{1}$, $\rho(s)$ is the spectral density, and $s_{0}$ is the threshold.
\par
On the OPE side, with the propagator of the heavy quark $Q$, formulated as
\begin{equation}
S(x')=-i\int\frac{d^{4}k}{(2\pi)^{4}}e^{-ik\cdot x'}\frac{\slashed{k}+m_{Q}}{m^{2}_{Q}-k^{2}},
\end{equation}
and the distribution amplitude (\ref{doda}), the correlation function on the theoretical side is expressed as
\begin{equation}
\Pi_{\mu}=2im_{Q}p_{\mu}\int^{1}_{0}dx\frac{\phi_{S}(x,\mu)}{m^{2}_{Q}-(q+xp)^{2}}.\\
\end{equation}
By matching the hadronic side with  the OPE side and applying the Borel transformation to both sides, we acquire the sum rules for the form factors \cite{Sun:2010nv}
\begin{equation}
\begin{aligned}
f_{+}(q^2)&=-\frac{m_{Q}+m_{q_{1}}}{m_{B/D}^{2}f_{B/D}}m_{Q}\int^{1}_{\Delta}\frac{\phi(x)}{x}dxe^{\mathcal{FF}},\\
f_{-}(q^2)&=\frac{m_{Q}+m_{q_{1}}}{m_{B/D}^{2}f_{B/D}}m_{Q}\int^{1}_{\Delta}\frac{\phi(x)}{x}dxe^{\mathcal{FF}},\\
\label{FF}
\end{aligned}
\end{equation}
where the function $\mathcal{FF}$ is given by
\begin{equation}
\begin{aligned}
\mathcal{FF}&=-\frac{1}{xM^{2}}(m^{2}_{Q}+x\bar{x}p^{2}-\bar{x}q^{2})+\frac{m^{2}_{M}}{M^{2}},
\end{aligned}
\end{equation}
and $\Delta$ is defined as
\begin{equation}
\begin{aligned}
\Delta&=\frac{\sqrt{(s_{0}-m_{S}^{2}-q^{2})^{2}+4(m_{Q}^{2}-q^{2})m_{S}^{2}}-(s_{0}-m_{S}^{2}-q^{2})}{2m_{S}^{2}}.
\end{aligned}
\end{equation}

\section{Numerical analyses and discussion}
The input parameters involved in the numerical calculation are as follows \cite{Huang:2004tp,ParticleDataGroup:2024cfk}
\par
\begin{centering}
$m_{u}=2.16\ \rm{MeV}  $,

$m_{d}=4.67\ \rm{MeV}  $,

$m_{a_{0}(980)}=0.98\ \rm{GeV}  $,

$m_{a_{0}(1450)}=1.474\ \rm{GeV}  $,

$\langle \bar{u}u\rangle=\langle \bar{d}d\rangle=-0.24^{3}\ \rm{GeV^{3}}  $,

$\langle g_{s}\bar{u}\sigma Gu\rangle=\langle g_{s}\bar{d}\sigma Gd\rangle=0.8\times\langle \bar{u}u\rangle=0.8\times\langle \bar{d}d\rangle  $,

$m_{0}^{2}=0.8\ \rm{GeV^{2}}  $,

$\alpha_{s}=0.517   $.

\end{centering}
\par
\begin{figure}[H]
\centering
\begin{minipage}[c]{0.5\textwidth}
\centering
\subfigure[$\xi^{1}$]{
\includegraphics[height=5cm,width=7cm]{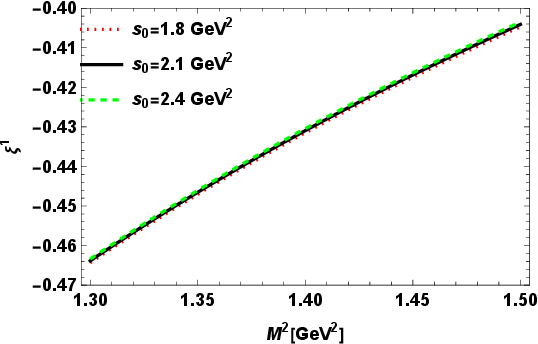}}
\end{minipage}%
\begin{minipage}[c]{0.5\textwidth}
\centering
\subfigure[$\xi^{3}$]{
\includegraphics[height=5cm,width=7cm]{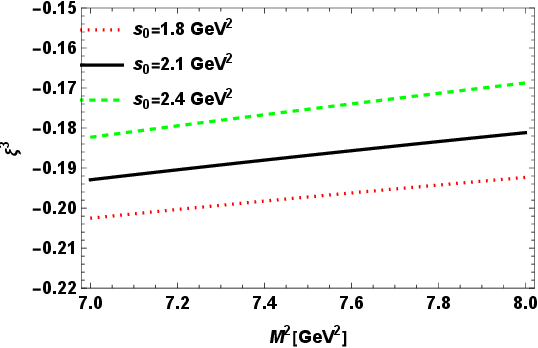}}
\end{minipage}
\caption{The first moment (left panel) $\xi^{1}$ and the third moment (right panel) $\xi^{3}$ of $a_{0}(980)$. The dot, solid and dash line correspond to threshold $s_{0}=1.8,\ 2.1,\ 2.4\ \rm{GeV^2}$ at the scale $\mu=1\ \rm{GeV}$.}
\label{fig.main-980}
\end{figure}
\begin{figure}[H]
\centering
\begin{minipage}[c]{0.5\textwidth}
\centering
\subfigure[$\xi^{1}$]{
\includegraphics[height=5cm,width=7cm]{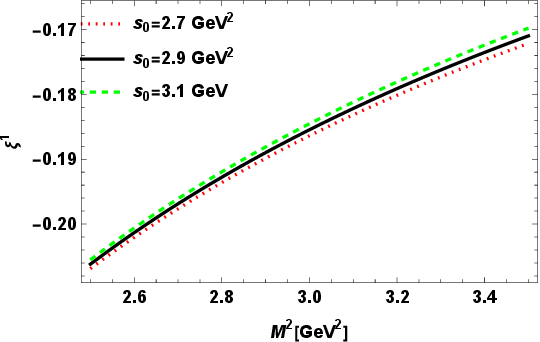}}
\end{minipage}%
\begin{minipage}[c]{0.5\textwidth}
\centering
\subfigure[$\xi^{3}$]{
\includegraphics[height=5cm,width=7cm]{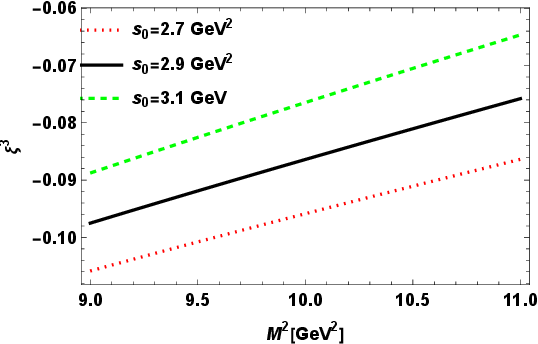}}
\end{minipage}
\caption{The first moment (left panel) $\xi^{1}$ and the third moment (right panel) $\xi^{3}$ of $a_{0}(1450)$. The dot, solid and dash line correspond to threshold $s_{0}=2.7,\ 2.9,\ 3.1\ \rm{GeV^2}$ at the scale $\mu=2\ \rm{GeV}$.}
\label{fig.main-1450}
\end{figure}

As for the parameters related to the meson $a_{0}(980)$, the scale is set at $\mu=1\ \rm{GeV}$, and for $a_{0}(1450)$, the scale is at $\mu=2\ \rm{GeV}$. The Borel parameter and threshold are chosen to be within ranges that adhere to the conventional constraints of sum rule calculations. To delineate the Borel window for $M^{2}$, we present an analysis of the Borel parameter curves for the moments in Figures \ref{fig.main-980} and \ref{fig.main-1450}. 
\par
Figure \ref{fig.main-980} displays the first moment ($\xi^{1}$, left panel) and the third moment ($\xi^{3}$, right panel) for $a_{0}(980)$. The Borel windows are determined as $[1.3, 1.5]\ \rm{GeV^{2}}$ for $\xi_{1}$ and $[7, 8]\ \rm{GeV^{2}}$ for $\xi_{3}$, with threshold parameters $s_{0}=1.8, 2.1, 2.4\ \rm{GeV^{2}}$. The dependence of $\xi^{n}$ on $M^{2}$ is negligible, while the threshold variation contributes less than $5\%$ to the moments, as indicated by the overlapping curves for different $s_{0}$.
\par
In an analogous manner, the Borel windows for the initial two moments of the $a_{0}(1450)$ meson have been determined and are presented in Figure~\ref{fig.main-1450}. The Borel window for the first moment is identified within the range of $[2.5, 3.5]\ \rm{GeV^{2}}$, while the third moment falls between $[9, 11]\ \rm{GeV^{2}}$. It is also observed that the influence of the threshold parameter on the moments derived from the QCD sum rule analysis is marginal.
\par
\begin{figure}[H]
\centering
\begin{minipage}[c]{0.5\textwidth}
\centering
\subfigure[$a_{0}(980)$]{
\includegraphics[height=5cm,width=7cm]{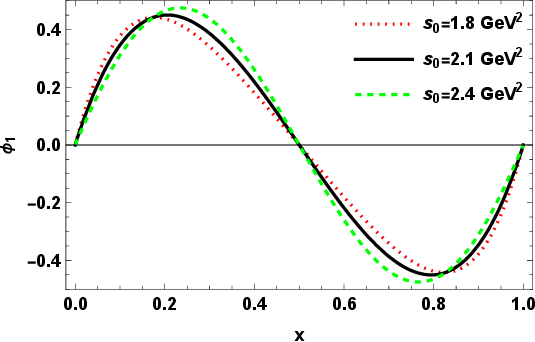}}
\end{minipage}%
\begin{minipage}[c]{0.5\textwidth}
\centering
\subfigure[$a_{0}(1450)$]{
\includegraphics[height=5cm,width=7cm]{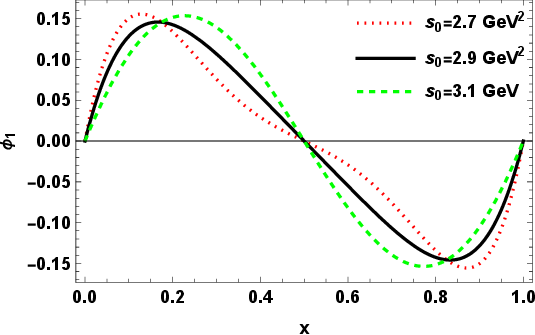}}
\end{minipage}
\caption{The distribution amplitudes of $a_{0}(980)$ (left panel) and $a_{0}(1450)$ (right panel). The dot, solid and dash line of left panel correspond to threshold $s_{0}=1.8,\ 2.1,\ 2.4\ \rm{GeV^2}$ at the scale $\mu=1\ \rm{GeV}$. The dot, solid and dash line of right panel correspond to threshold $s_{0}=2.7,\ 2.9,\ 3.1\ \rm{GeV^2}$ at the scale $\mu=2\ \rm{GeV}$.}
\label{fig.main-phi}
\end{figure}
\begin{figure}[H]
\centering
\begin{minipage}[c]{0.5\textwidth}
\centering
\subfigure[$a_{0}(980)$]{
\includegraphics[height=5cm,width=7cm]{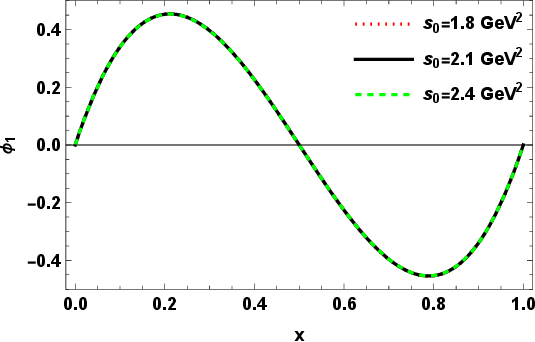}}
\end{minipage}%
\begin{minipage}[c]{0.5\textwidth}
\centering
\subfigure[$a_{0}(1450)$]{
\includegraphics[height=5cm,width=7cm]{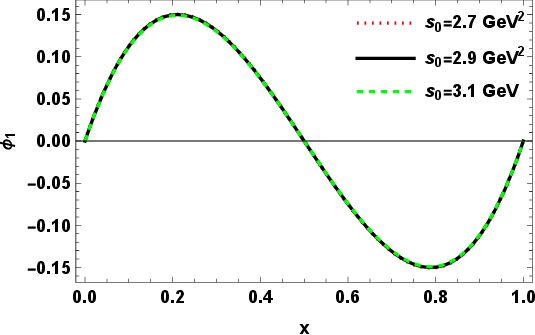}}
\end{minipage}
\caption{The distribution amplitudes truncated at the first Gegenbauer moment $\xi^{1}$ of $a_{0}(980)$ (left panel) and $a_{0}(1450)$ (right panel). The dot, solid and dash line of left panel correspond to threshold $s_{0}=1.8,\ 2.1,\ 2.4\ \rm{GeV^2}$ at the scale $\mu=1\ \rm{GeV}$. The dot, solid and dash line of right panel correspond to threshold $s_{0}=2.7,\ 2.9,\ 3.1\ \rm{GeV^2}$ at the scale $\mu=2\ \rm{GeV}$.}
\label{fig.main-phi-xi1}
\end{figure}

In detail, for the $a_{0}(980)$ meson, we obtain $\xi^{1} = -0.431 \pm 0.030 \pm 0.001$ and $\xi^{3} = -0.187 \pm 0.010 \pm 0.006$. The first uncertainty arises from the Borel parameter, while the second arises from the threshold. For the $a_{0}(1450)$ meson, the corresponding values are $\xi^{1} = -0.185 \pm 0.020 \pm 0.001$ and $\xi^{3} = -0.086 \pm 0.010 \pm 0.008$. According to Equation~(\ref{B-xi}), the Gegenbauer coefficients are presented in Table~\ref{B}.
\begin{table}[H]
\centering
\caption{Gegenbauer coefficients at the scale $\mu=1\ \rm{GeV}$ for $a_{0}(980)$ and $\mu=2\ \rm{GeV}$ for $a_{0}(1450)$.}
	\begin{tabular}{p{5cm}<{\centering}p{5cm}<{\centering}p{5cm}<{\centering}}
	\hline
        \hline
        ~&$B_{1}$&$B_{3}$\\
        \hline
        $a_{0}(980)$&-0.718&-0.011 \\
        \hline
        $a_{0}(1450)$&-0.308&-0.035\\
        \hline
        \hline
	\end{tabular}
\label{B}
\end{table}
\noindent Further, according to Equation~(\ref{phi-G}), one can obtain the distribution amplitudes of scalar mesons $a_{0}(980)$ and $a_{0}(1450)$ as shown in Figure \ref{fig.main-phi} and \ref{fig.main-phi-xi1}.
\par
Figure \ref{fig.main-phi} presents the twist-2 distribution amplitudes of $a_{0}(980)$ at the scale $\mu=1\ \rm{GeV}$ and $a_{0}(1450)$ at the scale $\mu=2\ \rm{GeV}$. To elucidate the reason of dependence of distribution amplitude on the threshold $s_{0}$, we show the distribution  amplitudes of $a_{0}(980)$ and $a_{0}(1450)$ with only the first moment in Figure \ref{fig.main-phi-xi1}. A clear $s_{0}$-independence is observed in these truncated calculations. This implies that the $s_{0}$-dependence of the full twist-2 distribution amplitudes for both $a_{0}(980)$ and $a_{0}(1450)$ stems exclusively from higher moments. Even when the third moment is incorporated into our analysis, it remains evident that the insensitivity of the distribution amplitude of $a_{0}(980)$ to $s_{0}$ contrasts with the sensitivity of that of $a_{0}(1450)$ to $s_{0}$ as shown in Figure \ref{fig.main-phi}. 
\par
It is well known that the distribution amplitude quantifies the longitudinal momentum fraction carried by valence quarks (or partons) within a meson in the infinite-momentum frame, encoding essential non-perturbative information about its light-cone wavefunction. For conventional quark-antiquark ($q\bar{q}$) configurations, the distribution amplitude of a scalar meson typically exhibits a double extremum, reflecting symmetric momentum between constituents. The $a_{0}(980)$ meson conforms to this scenario, with the extremum of the distribution amplitude near $x\sim0.2/0.8$. In contrast, exotic configurations (e.g., tetraquark $qq\bar{q}\bar{q}$ or molecular states) may manifest suppress endpoints ($x \to 0,1$) and enhance mid-momentum components, arising from additional color correlations or cluster substructures. The distribution amplitude of $a_{0}(1450)$ shows a broader distribution, suggesting possible tetraquark admixtures. This conclusion is predicated on the assumption that both states $a_{0}(980)$ and $a_{0}(1450)$ are treated as ground-state particles. However, it should be noted that the $a_{0}(1450)$ meson is conventionally regarded as the first excited state of the $a_{0}(980)$ meson rather than the ground state.

\par
Finally, by substituting the distribution amplitudes illustrated in Figure \ref{fig.main-phi} into Eq. (\ref{FF}) and using the Borel and threshold parameters from Ref. \cite{Huang:2021owr}, transition form factors of $B\rightarrow a_{0}(980)$, $B\rightarrow a_{0}(1450)$, $D\rightarrow a_{0}(980)$ and $D\rightarrow a_{0}(1450)$ are obtained. These results are presented in Table \ref{CFF} and compared with those of other methods.

\begin{table}[H]
\begin{center}
\caption{Comparison of form factors for $B/D\rightarrow a_{0}$ in this work with QCD sum rule (SR), light cone sum rule (LCSR), perturbative QCD (pQCD), Covariant Confining Quark Model (CCQM) and Covariant Light-Front quark Model (CLFQM).}
\begin{tabular}{p{7cm}<{\centering}p{1.5cm}<{\centering}p{1.5cm}<{\centering}p{1.5cm}<{\centering}p{1.5cm}<{\centering}p{1.5cm}<{\centering}p{1.5cm}<{\centering}p{1.5cm}<{\centering}p{1.5cm}<{\centering}p{1.5cm}<{\centering}}
\hline
\multicolumn{2}{c}{Process} & \multicolumn{2}{c}{$B\rightarrow a_{0}(980)$} & \multicolumn{2}{c}{$B\rightarrow a_{0}(1450)$} & \multicolumn{2}{c}{$D\rightarrow a_{0}(980)$} & \multicolumn{2}{c}{$D\rightarrow a_{0}(1450)$} \\
\hline
\multicolumn{2}{c}{Form factor}&$f_{+}$&$f_{-}$&$f_{+}$&$f_{-}$&$f_{+}$&$f_{-}$&$f_{+}$&$f_{-}$\\
\hline
\multicolumn{2}{c}{Our work}&0.53&-0.53&0.23&-0.23&0.77&-0.77&0.32&-0.32\\

\multicolumn{2}{c}{SR1 \cite{Sun:2010nv}}&0.56&-0.56&0.26&-0.26&~&~&~&~\\

\multicolumn{2}{c}{SR2 \cite{Sun:2010nv}}&~&~&0.53&-0.53&~&~&~&~\\

\multicolumn{2}{c}{LCSR \cite{Wang:2008da}}&~&~&0.52&-0.44&~&~&~&~\\

\multicolumn{2}{c}{pQCD \cite{Li:2008tk}}&0.39&~&-0.31&~&~&~&~&~\\

\multicolumn{2}{c}{pQCD \cite{Li:2008tk}}&~&~&0.68&~&~&~&~&~\\

\multicolumn{2}{c}{LCSR \cite{Huang:2021owr}}&~&~&~&~&0.85&-0.85&0.94&-0.94\\

\multicolumn{2}{c}{LCSR \cite{Cheng:2017fkw}}&~&~&~&~&1.75&0.31&~&~\\

\multicolumn{2}{c}{CCQM\cite{Soni:2020sgn}}&~&~&~&~&0.55&~&~&~\\

\multicolumn{2}{c}{CLFQM\cite{Verma:2011yw}}&~&~&~&~&~&~&0.51&~\\
\hline
\end{tabular}
\label{CFF}
\end{center}
\end{table}
\par
For the form factors for the process $D^{+}\rightarrow a_{0}^{0}(980)e^{+}\nu_{e}$, our predictions of branching fraction 0.79 exhibits agreement with BES\uppercase\expandafter{\romannumeral3} measurement \cite{BESIII:2018sjg} $1.66^{+0.81}_{-0.66}$. This suggests the dominance of the $q\bar{q}$ component in the $a_{0}(980)$. However, significant discrepancies arise in the form factors related to $a_{0}(1450)$ when comparing our results with other theoretical studies. These discrepancies suggest that the experimentally observed $a_{0}(1450)$ resonance may not strictly be a single configuration. Instead, it could manifest as an admixture of two distinct components: (i) an intrinsic ground-state $a_{0}(1450)$ in the conventional quark-antiquark picture, and (ii) the first excited state of $a_{0}(980)$ arising from radial excitation.
\par
Furthermore, a more nuanced interpretation must consider the non-trivial structure of these states. Specifically, the mesons $a_{0}(980)$ and $a_{0}(1450)$ may not be pure $q\bar{q}$ states. Alternative configurations, such as tetraquark ($qq\bar{q}\bar{q}$) states or quantum mechanical superpositions of conventional and multiquark components, remain plausible and could contribute to the observed phenomenological behavior.

\section{Summary} 
We present a systematic QCD sum rule analysis of the twist-2 distribution amplitudes for the $a_0(980)$ and $a_0(1450)$ mesons. By constructing appropriate correlation functions and calculating the first two non-zero moments $\langle \xi^1 \rangle$ and $\langle \xi^3 \rangle$, we determine the Gegenbauer coefficients $B_1$ and $B_3$ for these mesons. Our results reveal distinct dynamical features: the $a_0(980)$ meson exhibits significant asymmetry in its distribution amplitude ($\xi^1 = -0.431$, $\xi^3 = -0.187$), with the extremum occurring near $x\sim 0.2/0.8$, consistent with a conventional $q\bar{q}$ configuration. In contrast, the $a_0(1450)$ meson displays a broader distribution amplitude ($\xi^1 = -0.185$, $\xi^3 = -0.086$) suggesting that $a_0(1450)$ may not strictly be a single configuration when $a_0(1450)$ is treated as a ground-state particle.
\par
We further compute the transition form factors for $B/D \to a_0$ decays using light-cone sum rules. Predictions for the transition form factors $f_{+}$ and $f_{-}$ show agreement with experimental data for $a_0(980)$ but highlight discrepancies for the form factors for $a_0(1450)$, implying a complex structure for the latter. These results provide critical inputs for heavy-flavor phenomenology and underscore the necessity of incorporating higher-twist effects. Our work advances the understanding of scalar meson structures and offers valuable information for experimental tests at facilities like LHCb and Belle \uppercase\expandafter{\romannumeral2}.

\section*{ACKNOWLEDGEMENTS}
Y. J. Sun is supported in part by the National Natural Science Foundation of China under the Grant No. 11365018 and No. 11375240.
\par

\makeatletter\def\@captype{table}\makeatother

\end{document}